\documentclass[reprint,aps,physrev,twocolumn,showpacs,showkeys, 10pt, longbibliography, floatfix, amsmath,amssymb, nolinenumbers
]{revtex4-2}

\usepackage{graphicx}
\usepackage{dcolumn}
\usepackage{bm}
\usepackage{amsfonts}

\usepackage[colorlinks = true,linkcolor = blue,urlcolor  = blue,citecolor = blue,anchorcolor = blue]{hyperref}
\usepackage{float}
\usepackage{times}
\usepackage{orcidlink}
\usepackage{xcolor}
\usepackage{multirow}
\usepackage[normalem]{ulem}

\usepackage[
  protrusion=true,
  expansion=true,
  tracking=true,
  kerning=true,
  spacing=true
]{microtype}

\begin{document}

\preprint{APS/123-QED}

\title{Impact of Kaon Condensation on the Thermal Evolution of the CCO in HESS J1731--347 Supernova Remnant
}%

\author{D.G. Nanopoulos\orcidlink{0009-0003-4349-6392}$^{1}$}
\email{dnanopou@physics.auth.gr}
\author{P.S. Koliogiannis\orcidlink{0000-0001-9326-7481}$^{2}$}
\email{pkoliogi@phy.hr}
\author{V. Petousis\orcidlink{0000-0002-5575-6476}$^{3}$}
\email{vlasios.petousis@cvut.cz}
\author{M. Veselsk\'y\orcidlink{0000-0002-7803-0109}$^{3}$}
\email{Martin.Veselsky@cvut.cz}
\author{Ch.C. Moustakidis\orcidlink{0000-0003-3380-5131}$^{1}$}
\email{moustaki@auth.gr}

\affiliation{$^{1}$Department of Theoretical Physics, Aristotle University of Thessaloniki, 54124 Thessaloniki, Greece}
\affiliation{$^{2}$Department of Physics, Faculty of Science, University of Zagreb, Bijeni\v cka cesta 32, 10000, Zagreb, Croatia}
\affiliation{$^{3}$Institute of Experimental and Applied Physics, Czech Technical University, 110 00, Prague, Czechia}

\date{\today}

\begin{abstract}
Recent analyses of the central compact object in the HESS J1731--347 supernova remnant suggest an unusual combination of a low mass and small radius, while its thermal emission indicates a relatively high surface temperature at its estimated age. Reconciling these structural and thermal properties within a unified theoretical framework may provide important constraints on the equation of state and composition of dense matter. In this work, we investigate the thermal consequences of negatively charged kaon condensation, an exotic phase that softens the equation of state and facilitate the reproduction of the inferred bulk properties of HESS J1731--347. We find that the onset of kaon condensation strongly accelerates the thermal evolution, leading to surface temperature substantially below the observationally inferred range. Within the adopted cooling framework, kaon condensation therefore cannot simultaneously account for the structural and thermal properties of HESS J1731--347.
\end{abstract}

\maketitle


\section{Introduction}
\label{introduction}
Neutron-star cooling is a complex process governed by the composition and underlying equation of state (EoS) of the stellar interior, together with microscopic properties such as thermal conductivity, superfluidity and superconductivity~\cite{Flowers-1981, Baiko-1999, Tamagaki-1970, Alford-2001a, Alford-2001b, Alford-2008}. Since several of these ingredients remain uncertain, comparing the predicted thermal evolution with observationally inferred ages and surface temperatures provides a valuable
means of probing dense matter. When combined with mass and radius
measurements, thermal observations can further constrain its composition and EoS~\cite{Klochkov-2015}.

Doroshenko et al.~\cite{Doroshenko-2022} analysed the central compact object (CCO) in the HESS J1731--347 supernova remnant and inferred a mass of $M=0.77^{+0.20}_{-0.17}~M_\odot$ and a radius of
$R=10.4^{+0.86}_{-0.78}~\mathrm{km}$. These values place the CCO among the lightest and smallest compact stars observed to
date. The combination of such a low mass and small radius is unusual
within the conventional picture of neutron stars (NSs)~\cite{Zhang-2011} and has motivated several interpretations of its nature. In particular, the CCO has been studied as a hadronic star~\cite{Klochkov-2015, Zhang-2025,Brodie-2023,Huang-2024,Li-2023,Kubis-2023,Char-2026,Tewari-2025}, a hybrid star~\cite{Tsaloukidis-2023,Sagun-2023,Laskos-2024,Laskos-2025,Li-2024,Mariani-2024,Gholami-2025,Gao-2024,Pal-2025,Alvarez-2025,Kubis-2025}, a strange star~\cite{Di-Clemente-2024,Horvath-2023,Oikonomou-2023,Das-2023,Rather-2023,Kourmpetis-2025}, a dark-matter-admixed star~\cite{Yang-2024}, and a kaon-condensed star~\cite{Veselsky-2025a,Veselsky-2025b}.

Its thermal properties provide an additional constraint on these
interpretations, where a redshifted surface temperature of $T_s^\infty = 2.05 ^{+0.09} _{-0.06}$ MK has been reported~\cite{Doroshenko-2022, Sagun-2023}. Combined with the initial age estimate of $\sim 27~\mathrm{kyr}$, this temperature appeared difficult
to explain~\cite{Horvath-2023}. More recent studies, however, favour a younger age in the range of $2$--$6~\mathrm{kyr}$~\cite{Horvath-2023,Acero-2015,Cui-2016,Maxted-2018}. Di Clemente et al.~\cite{Di-Clemente-2024} and Horvath et al.~\cite{Horvath-2023} suggested that color-superconducting strange quark matter could account for the thermal properties of the CCO, although without performing detailed cooling simulations. Yuan et al.~\cite{Yuan-2025} subsequently studied its thermal evolution as a color-flavor-locked strange star and showed that an additional heating mechanism, such as $r$-mode heating, could reproduce its thermal properties. Sagun et al.~\cite{Sagun-2023}
considered both hadronic and hybrid configurations, using the initial age estimate, and found that agreement with the surface-temperature constraint required paired stellar matter. Using the revised age range, Zhang et al.~\cite{Zhang-2025} similarly emphasized the role of nucleon superfluidity in the hadronic-star interpretation. More recently, Nanopoulos et al.~\cite{Nanopoulos-2026} extended the cooling analyses to hadronic, hybrid, and strange-star configurations. They found that the observational constraints can be reproduced within all three scenarios, although the exotic configurations require quark pairing and the suppression of efficient direct Urca emission.

In this work, we investigate whether kaon condensation can account simultaneously for the structural and thermal properties of HESS J1731--347. Although kaon-condensed EoSs have recently been shown to reproduce its unusually low mass and small radius~\cite{Veselsky-2025a,Veselsky-2025b}, their compatibility with the
observed surface temperature remains unclear. Within the same nuclear framework developed in Ref.~\cite{Veselsky-2025a}, which has been successfully applied in various extensions and modifications to the kaon-condensed matter, we calculate the thermal evolution of kaon-condensed stars by including the relevant neutrino-emission processes and nucleon-pairing effects, and compare the resulting cooling curves with the temperature and age inferred for the CCO.

The paper is structured as follows. Section~\ref{sec:Theoretical_Framework} presents the theoretical framework underlying the EoSs, with particular emphasis on kaon condensation and the cooling processes. Section~\ref{sec:Results} presents the results and discusses their implications. Finally, Section~\ref{sec:Conclusion} summarizes the main conclusions.

\section{Theoretical Framework}
\label{sec:Theoretical_Framework}
\subsection{Matter in a \texorpdfstring{$\beta$}{beta}-equilibrium state}
In the present work, neutron-star matter is assumed to consist of neutrons, protons, electrons, and negatively charged kaons, while neutrinos are neglected under the assumption of cold catalyzed matter. The composition of matter is determined by imposing chemical equilibrium among the constituent particles through weak interaction processes.

For nucleonic matter, chemical equilibrium is established through the $\beta$-decay and inverse $\beta$-decay reactions~\cite{Thorsson-1994,Lim-2014}
\begin{equation}
    n \rightarrow p + e^{-} + \bar{\nu}_{e},
    \quad
    p + e^{-} \rightarrow n + \nu_{e},
    \label{eq:beta_proc}
\end{equation}
which, after neutrinos escape from the system, lead to the equilibrium condition
\begin{equation}
    \mu_n-\mu_p=\mu_e,
    \label{eq:chem_eq}
\end{equation}
where $\mu_n$, $\mu_p$, and $\mu_e$ denote the chemical potentials of neutrons, protons, and electrons, respectively.

When the onset condition for kaon condensation is satisfied, the relevant weak interaction processes become
\begin{equation}
    n \leftrightarrow p + K^{-},
    \quad
    e^{-} \leftrightarrow K^{-} + \nu_{e},
    \label{eq:strangeness_proc}
\end{equation}
leading to the equilibrium conditions
\begin{equation}
    \mu_n-\mu_p=\mu_{K^{-}},
    \quad
    \mu_e=\mu_{K^{-}}\equiv\mu,
    \label{eq:strangeness_chem_eq}
\end{equation}
where $\mu_{K^{-}}$ is the chemical potential of the negatively charged kaon condensate.

\subsection{Equation of State}
The EoSs adopted in the present work were obtained within the chiral effective model coupled to the momentum-dependent interaction (MDI) framework developed in Ref.~\cite{Veselsky-2025a, Thorsson-1994}. In this approach, the strength of the kaon--nucleon interaction is determined by the parameters $a_{1}m_{s}$, $a_{2}m_{s}$, and $a_{3}m_{s}$, with $a_{1}m_{s}=-67~\mathrm{MeV}$ and $a_{2}m_{s}=134~\mathrm{MeV}$. The parameter $a_{3}m_{s}$, which is directly related to the strangeness content of the proton, is treated as a free parameter within the range $[-134,-310]~\mathrm{MeV}$, corresponding to proton strangeness contents between $0\%$ and $20\%$, respectively. Variations of $a_{3}m_{s}$ modify the onset density and abundance of the kaon condensate and consequently may affect the neutrino emissivity and cooling evolution of NSs. In following, the final expressions for the energy density and pressure are summarized, while the complete derivation can be found in Ref.~\cite{Veselsky-2025a}.

The energy density is given by
\begin{align}
    \mathcal{E}(u,x,\mu,\theta)
    &={}
    \mathcal{E}_{\mathrm{MDI}}\left(u,x=\tfrac{1}{2}\right)
    +un_0(1-2x)^2S(u) \nonumber
    \\
    &+
    \frac{f^2\mu^2}{2(\hbar c)^3}\sin^2\theta
    +
    \frac{2f^2m_K^2c^4}{(\hbar c)^3}
    \sin^2\left(\frac{\theta}{2}\right) \nonumber
    \\
    &+
    \left(2a_1m_sx+\mathcal{T}_{23}\right)
    un_0\sin^2\left(\frac{\theta}{2}\right)
    +\mathcal{E}_e,
    \label{eq:energy_density}
\end{align}
where ${\cal E}_{\rm MDI}$ and $S(u)$ correspond to the energy density and symmetry energy of the MDI+APR1 EoS~\cite{Koliogiannis-2021}, ${\cal E}_{e}$ is the electron energy density
\begin{equation}
    {\cal E}_{e}
    =
    \frac{\mu^{4}}
    {4\pi^{2}(\hbar c)^3},
    \label{eq:energy_density_el}
\end{equation}
with $f=93~\mathrm{MeV}$ the pion decay constant, $m_{K}$ the kaon mass, $n_{0} = 0.16~\mathrm{fm}^{-3}$ the nuclear saturation density, $u=n/n_{0}$ the normalized baryon density, $x$ the proton fraction, $\theta$ the kaon condensate amplitude, and $\mathcal{T}_{23}=2a_{2}m_{s}+4a_{3}m_{s}$.

The pressure is taken as the sum of the contributions from baryons, kaons, and electrons, namely
\begin{equation}
    p(u,x,\mu,\theta)
    =
    p_{b}(u,x)
    +
    p_{K}(\mu,\theta)
    +
    p_{e},
    \label{eq:pressure}
\end{equation}
where
\begin{align}
p_b(u,x)
&=
u^2\frac{\partial}{\partial u}
\left[\frac{\mathcal{E}_b(u,x)}{u}\right],
\\
p_K(\mu,\theta)
&=
-\frac{f^2\mu^2}{2(\hbar c)^3}\sin^2\theta
-\frac{2f^2m_K^2c^4}{(\hbar c)^3}
\sin^2\left(\frac{\theta}{2}\right),
\\
p_e(\mu)
&=
\frac{\mu^4}{12\pi^2(\hbar c)^3}.
\label{eq:pressure_comp}
\end{align}

In the present work, we consider three representative EoSs: the purely hadronic MDI+APR1 model and its two kaon-condensed variants with $a_{3}m_s=-238~\mathrm{MeV}$ (MDI+APR1-KC1) and $a_{3}m_s=-260~\mathrm{MeV}$ (MDI+APR1-KC2)~\cite{Veselsky-2025a}.

\subsection{Cooling processes in the core}
The most powerful neutrino emitting process in hadronic matter is the direct Urca (dURCA) process and it consists of the reactions given in Eq.~\eqref{eq:beta_proc}. In this case, the emissivity is~\cite{Yakovlev-1999, Yakovlev-2001}
\begin{equation} 
\label{eq:241}
    \mathcal{E}_\nu^{D} = \frac{457\pi}{10080} \mathcal{G}_{1} (f_V^2 + 3g_A^2) 
    \frac{m_n^* m_p^* m_e^*}{\hbar^{10} c^3} (k_BT)^6,
\end{equation}
where $\mathcal{G}_{1}=G_F^2 \cos^2{\theta_C}$, with $G_F = 1.436\times10^{-49}$ $\mathrm{erg~cm^3}$ being the Fermi constant of weak interaction and $\theta_C = 0.223$ the Cabibbo angle~\cite{Thorsson-1995}, $f_V = 1$ is the vector constant for the dURCA process, $g_A = 1.26$ the axial-vector constant for the specific reaction, and $m_{i=n,p,e}^*$ corresponds to the particle's effective mass. In the present work, we consider the neutron and proton effective masses to be $m_{j=n,p}^*=0.7m_{j=n,p}$~\cite{Yakovlev-1999,Yakovlev-2001,Ofengeim-2017}, whereas the electron effective mass is given by
$m_e^*=\mu_e/c^2\approx p_{F_e}/c$~\cite{Yakovlev-1999,Yakovlev-2001}, where $p_{F_e}$ denotes the electron Fermi momentum. 

When the dURCA process is not activated, the neutrino emission of hadronic matter is determined by the modified Urca (mURCA) process and the neutrino nucleon-nucleon Bremsstrahlung process. In particular, the mURCA process consists of two branches,
\begin{align}
    &
    \begin{array}[c]{@{}r@{\quad}l@{}}
    \text{Neutron branch:} &
    \left\{
    \begin{array}{@{}r@{\;\rightarrow\;}l@{}}
    n+n     & p+n+e^-+\bar{\nu}_e, \\
    p+n+e^- & n+n+\nu_e,
    \end{array}
    \right.
    \\[0.4ex]
    \text{Proton branch:} &
    \left\{
    \begin{array}{@{}r@{\;\rightarrow\;}l@{}}
    n+p     & p+p+e^-+\bar{\nu}_e, \\
    p+p+e^- & n+p+\nu_e,
    \end{array}
    \right.
    \end{array}
    &&
    \label{eq:murca}
\end{align}
and their respective emission rates defined as~\cite{Yakovlev-2001}
\begin{align}
    \mathcal{E}_\nu^{Mn} &= \mathcal{G}_{2} m_n^{*3}m_p^* \left(\frac{f^{\pi}}{m_{\pi}}\right)^4 p_{F_p}(k_BT)^8 \alpha_n \beta_n, \label{eq:242} \\
    \mathcal{E}_\nu^{Mp} &= \mathcal{E}_\nu^{Mn} \left(\frac{m_p^*}{m_n^*}\right)^2 \frac{(p_{F_e} + 3p_{F_p} - p_{F_n})^2}{8p_{F_e}p_{F_p}}, \label{eq:243}
\end{align}
where 
\begin{equation}
    \mathcal{G}_{2} = \frac{11513}{30240}\frac{\mathcal{G}_{1} g_A^2}{2\pi \hbar^{10}c^8},
\end{equation}
$p_{F_{n}}$ and $p_{F_{p}}$ are the neutron and proton Fermi momenta, respectively, $f^\pi \approx 1$ is the $\pi$N-interaction constant in the p-state in the one-pion-exchange approximation, and $m_\pi$ is the pion mass ($\pi^\pm$). Following Ref.~\cite{Yakovlev-2001}, the factors $\alpha_n$ and $\beta_n$ were set equal to 1.13 and 0.68, respectively. It is noted that the proton branch is allowed only when the inequality $p_{F_n} < 3p_{F_p} + p_{F_e}$ holds~\cite{Yakovlev-1999, Yakovlev-2001}. The emissivity of the neutron branch was initially calculated by Friman and Maxwell~\cite{Friman-1979} and of the proton branch by Yakovlev and Levenfish~\cite{Yakovlev-1995}.
 
The neutrino Bremsstrahlung reactions considered here can be written in the form
\begin{equation}
    N + N \rightarrow N + N + \nu + \bar{\nu},
    \label{eq:neutrino_brem}
\end{equation}
where $N\in\{n,p\}$ and $\nu$ denotes a neutrino of any flavour. The corresponding neutrino emissivities due to nucleon-nucleon scattering are given by~\cite{Yakovlev-2001}
\begin{align}
    \mathcal{E}^{nn}_\nu &= \mathcal{G}_{3} m_n^{*4} \left(\frac{f^{\pi}}{m_{\pi^{0}}}\right)^4 p_{F_n} \alpha_{nn} \beta_{nn} (k_BT)^8  N_\nu,  \label{eq:244} \\
    \mathcal{E}^{np}_\nu &= 2\mathcal{G}_{3} m_n^{*2}m_p^{*2} \left(\frac{f^{\pi}}{m_{\pi^{0}}}\right)^4 p_{F_p} \alpha_{np} \beta_{np} (k_BT)^8  N_\nu, \label{eq:245} \\
    \mathcal{E}^{pp}_\nu &= \mathcal{G}_{3} m_p^{*4} \left(\frac{f^{\pi}}{m_{\pi^{0}}}\right)^4 p_{F_p} \alpha_{pp} \beta_{pp} (k_BT)^8  N_\nu,  \label{eq:246}
\end{align}
where 
\begin{equation}
\mathcal{G}_{3} = \frac{41}{14175}  \frac{G_F^2g_A^2} {2\pi\hbar^{10}c^8},    
\end{equation}
and $N_\nu = 3$ is the number of neutrino flavours. Following Ref.~\cite{Yakovlev-2001}, we adopt
$(\alpha_{nn},\alpha_{np},\alpha_{pp})=(0.59,1.06,0.11)$,
together with $(\beta_{nn},\beta_{np})=(0.56,0.66)$ and
$\beta_{pp}\approx0.7$.

The onset of kaon condensation introduces a reduction factor in the dURCA emissivity and opens two additional reaction channels, known as the kaon-induced Urca (kURCA) processes~\cite{Thorsson-1995}:
\begin{align}
    &
    \begin{array}[c]{@{}r@{\quad}l@{}}
    \text{n-kURCA:} &
    \left\{
    \begin{array}{@{}r@{\;\rightarrow\;}l@{}}
    n(K)     & n(K)+e^-+\bar{\nu}_e, \\
    n(K)+e^- & n(K)+\nu_e,
    \end{array}
    \right.
    \end{array}
    &&
    \label{eq:nkurca}
    \\[0.4ex]
    &
    \begin{array}[c]{@{}r@{\quad}l@{}}
    \text{p-kURCA:} &
    \left\{
    \begin{array}{@{}r@{\;\rightarrow\;}l@{}}
    p(K)     & p(K)+e^-+\bar{\nu}_e, \\
    p(K)+e^- & p(K)+\nu_e,
    \end{array}
    \right.
    \end{array}
    &&
    \label{eq:pkurca}
\end{align}
where $n(K)$, $p(K)$ denote an excitation which is a superposition of a neutron and a proton in the presence of a kaon condensate. The emission rates and conditions, which should be satisfied for the activation of Eqs.~\eqref{eq:nkurca} and~\eqref{eq:pkurca}, are summarized in Table~\ref{tab:1}.

\begin{table}
    \centering
    \caption{Neutrino emissivities and corresponding conditions for the processes considered in the kaon-condensed phase~\cite{Thorsson-1994,Thorsson-1995,Kubis-2003,Kubis-2006}.}
    \label{tab:1}
    \begin{ruledtabular}
        \begin{tabular}{lll}
            Cycle & $\mathcal{E}/\mathcal{E}_{\nu}^D$ & Condition \\
            \hline
            dURCA & $\cos^2{(\theta/2)}$ & $|p_{F_p} - p_{F_n}| < p_{F_e} < p_{F_p} + p_{F_n}$ \\
            n-kURCA & $1/4\sin^2{\theta}\tan^2{\theta_C}$ & $2p_{F_n}>p_{F_e}$ \\
            p-kURCA & $\sin^2{\theta}\tan^2{\theta_C}$ & $2p_{F_p}>p_{F_e}$ \\
        \end{tabular}
    \end{ruledtabular}
\end{table}

\subsection{Specific heat in the core}
The total specific heat of the core in hadronic matter is given by $c_{tot}=c_n + c_p + c_e$, where the neutron, proton, and electron contributions are~\cite{Yakovlev-1999, Ofengeim-2017, Grigorian-2005}
\begin{align}
    c_j &= \frac{k_B^2}{3\hbar^3} T m_j^* p_{F_j}, \quad j = n, p & \label{eq:251} \\
    c_e &= 0.6\times10^{20} \left(\frac{n_e}{n_0}\right)^{2/3} T_9 \quad \mathrm{(erg~cm^{-3}~K^{-1})}, & \label{eq:252}
\end{align}
where $n_e$ is the electron number density and $T_9=T/(10^9~\mathrm{K})$. The contribution of kaons to $c_{tot}$ is negligible~\cite{Bhat-2026}.

\subsection{Pairing in the core}
Baryon superfluidity plays a crucial role in the thermal evolution of a compact object. In the present work, we use the Bardeen-Cooper-Schrieffer model of baryon superfluidity. More specifically, we consider protons and neutrons of the star's core to undergo singlet- ($^1S_0$) and triplet-state ($^3P_2$) pairing, respectively~\cite{Yakovlev-2004, Baldo-1990, Baldo-1992, Chen-1993, Takatsuka-1993, Elgaroy-1996a, Amundsen-1985a, Schaab-1996}. In these states, the critical temperature is related to the superfluid energy gap at zero temperature, $\Delta_{0_{jj}} \equiv \Delta_{jj}(0)$, as~\cite{Yakovlev-1999, Yakovlev-2001, Kaminker-2002, Andersson-2005}
\begin{align}
    &   \text{$^1S_0$ state:}\qquad  k_B T_{c_p} \simeq 0.5669 \Delta_{0_{pp}}, &
    \\[1ex]
    &   \text{$^3P_2$ state:}\qquad  k_B T_{c_n} \simeq 0.8416 \Delta_{0_{nn}}. &
\end{align}
Regarding the zero-temperature energy gap, we adopt the formula given in Refs.~\cite{Nanopoulos-2026,Andersson-2005,Ho-2015}, which is similar to the one in Refs.~\cite{Kaminker-2002,Kaminker-2001}
\begin{equation} 
\label{eq:261}
    \Delta_{0_{jj}}(k_{F_j}) = \Delta_0 \frac{(k_{F_j}-k_0)^2}{(k_{F_j}-k_0)^2+k_1} \frac{(k_{F_j}-k_2)^2}{(k_{F_j}-k_2)^2+k_3},
\end{equation}
where $k_{F_j}$ is the nucleon Fermi wavenumber~\cite{Kaminker-2002, Kaminker-2001}, and $\Delta_0$, $k_{0-3}$ are parameters determined by different superfluidity gap models; their values are taken from Table~1 of Ref.~\cite{Andersson-2005}. Eq.~\eqref{eq:261} is valid when $k_0<k_{F_j}<k_2$ holds~\cite{Kaminker-2002, Andersson-2005, Kaminker-2001}. The interpolation formula between the zero-temperature gap and the energy gap with respect to temperature is of the form $\Delta_{jj}(T) = \Delta_{0_{jj}} \sqrt{1-T/T_{c_j}}$ (see Ref. \cite{Grigorian-2005} and references therein).  

Below their respective critical temperatures $T_{c_j}$, neutron and proton pairing suppresses both the nucleon specific heats and the emissivities of the relevant neutrino processes. We account for this effect by multiplying the corresponding normal-state expressions by dimensionless reduction factors, which are related to the superfluid energy gap as $\xi_{jj} = e^{-\Delta_{jj}/k_BT}$. The specific combinations of these factors entering each neutrino process and the neutron and proton specific heats are summarized in Ref.~\cite{Nanopoulos-2026}. In the kaon-condensed phase, the n-kURCA and p-kURCA channels each involve a single nucleon species and are therefore suppressed by the pairing gap of that species~\cite{Grigorian-2005}. Following Ref.~\cite{Bhat-2026}, the n-kURCA and p-kURCA emissivities are thus multiplied by $\xi_{nn}$ and $\xi_{pp}$, respectively.

The onset of baryon superfluidity in the stellar interior activates two additional neutrino-emission channels associated with Cooper-pair breaking and formation: nPBF and pPBF for neutrons and protons, respectively~\cite{Schaab-1996}. The corresponding emissivities for $T<T_{c_j}$ are given by~\cite{Blaschke-2004, Voskresensky-1987}
\begin{align}
    \mathcal{E}_{\nu}^{j\mathrm{PBF}}
    &\sim 10^{29}\,m_{j\mathrm{PBF}}^*
    \left[\frac{p_{F_j}(n_b)}{p_{F_n}(n_0)}\right]
    \left[6.24\cdot10^5\frac{\Delta_{jj}}{\mathrm{erg}}\right]^7 \nonumber
    \\
    &\times
    \left(\frac{k_BT}{\Delta_{jj}}\right)^{1/2}
    \xi_{jj}^{\,2}\qquad
    (\mathrm{erg~cm^{-3}~s^{-1}}),
    \label{eq:262}
\end{align}
where $m_{jPBF} = m_j^*/m_N$, with $m_N$ being the nucleon mass.

\subsection{Crust and envelope}
In the crust, neutrino emission is enhanced by the Bremsstrahlung of electrons which scatter off atomic nuclei \cite{Yakovlev-1999}. Following Refs.~\cite{Yakovlev-1999, Blaschke-2001}, the crustal contribution to the total specific heat is neglected. The approximate neutrino luminosity associated with electron-nucleus bremsstrahlung was introduced by Maxwell~\cite{Maxwell-1979} and is given by
\begin{equation} 
\label{eq:271}
    L_{br} = 1.65\times 10^{39}\:\frac{M_{cr}}{M_\odot}\left(\frac{T_b}{10^9K}\right)^6e^{\nu_b} \quad (\mathrm{erg~s^{-1}}),
\end{equation}
where $M_{cr}$ is the crustal mass, $T_b$ is the temperature at the crust-envelope transition point, and $\nu_b$ is the value of the gravitational redshift at that point. 

The envelope, which acts as a thermal insulator, affects the surface temperature of the compact object. The relation between the internal temperature in the crust-envelope boundary, $T_b$, and the surface temperature, $T_s$, depends on the envelope's chemical composition. In the present cooling simulations, we have considered two envelope compositions: (a) a heavy-element (iron-like) envelope and (b) a light-element (helium-like) envelope. The corresponding relations are~\cite{Gudmundsson-1983, Cumming-2017}
\begin{align}
    &   \text{Fe-envelope:}\qquad  T_{b_8} = 1.288\:(T_{s_6}^4/g_{s_{14}})^{0.455},  \label{eq:272} &
    \\[1ex]
    &   \text{He-envelope:}\qquad  T_{b_8} = 0.552\:(T_{s_6}^4/g_{s_{14}})^{0.413}, \label{eq:273} &
\end{align}
with $g_{s_{14}} = GM(10^{14}R^2)^{-1} [1 - 2GM/(Rc^2)]^{-1/2}$, and $M$, $R$ the mass and radius of the compact star, respectively.

\subsection{Cooling in the isothermal approximation}
The thermal evolution of a spherically symmetric star is described by the general relativistic equations derived by Thorne~\cite{Thorne-1977}. In the present work, we considered an isothermal stellar interior, which dictates that the redshifted internal temperature $T^\infty(t) = T(r,t)e^{\nu(r)/2}$ is constant, where $e^{\nu(r)/2}$ is the time-time component of the metric tensor. As a result, the cooling of a compact star was reduced into the global thermal balance equations~\cite{Yakovlev-2001, Yakovlev-2004, Yakovlev-2011, Page-2006} extracted by Glen and Sutherland~\cite{Glen-1980}
\begin{align}
    C(T^\infty)\frac{dT^\infty}{dt} &= -L_{\nu}^\infty(T^\infty) - L_{\gamma}^\infty(T_s),  \label{eq:281} \\
    C &= \sum_{i}\int c_{vi}dV, \label{eq:282} \\
    L_{\nu}^\infty &= \sum_{i}\int \mathcal{E}_ie^{\nu(r)}dV, \label{eq:283} \\
    L_{\gamma}^\infty &= 4\pi  R^2  \sigma  T_s^4  e^{\nu(r_b)}, \label{eq:284} \\
    dV &= 4 \pi r^2 \left[1 - \frac{2 G m(r)}{rc^2}\right]^{-1/2}dr, \label{eq:285}
\end{align}
where $C$ is the total heat capacity of the star, $L_\nu^\infty$ is the total neutrino luminosity detected by a distant observer, $L_\gamma^\infty$ is the redshifted photon luminosity, $\sigma$ is the Stefan-Boltzmann constant, $m(r)$ is the gravitational mass distribution of the star, and $r_b$ corresponds to the crust-envelope transition radius. 

Since the envelope is assumed to be extremely thin, the gravitational redshift factor at its base can be approximated by its surface value~\cite{Yakovlev-2011},
\begin{equation}
    e^{\nu(r_b)/2} \simeq e^{\nu(R)/2} = \sqrt{1-2GM/(Rc^2)}.
\end{equation}
It has to be noted that the isothermal approximation becomes valid once the star has undergone thermal relaxation, typically at ages $t \geq 10-10^3$yrs~\cite{Yakovlev-1999}.

\begin{figure}
\includegraphics[width=\columnwidth]{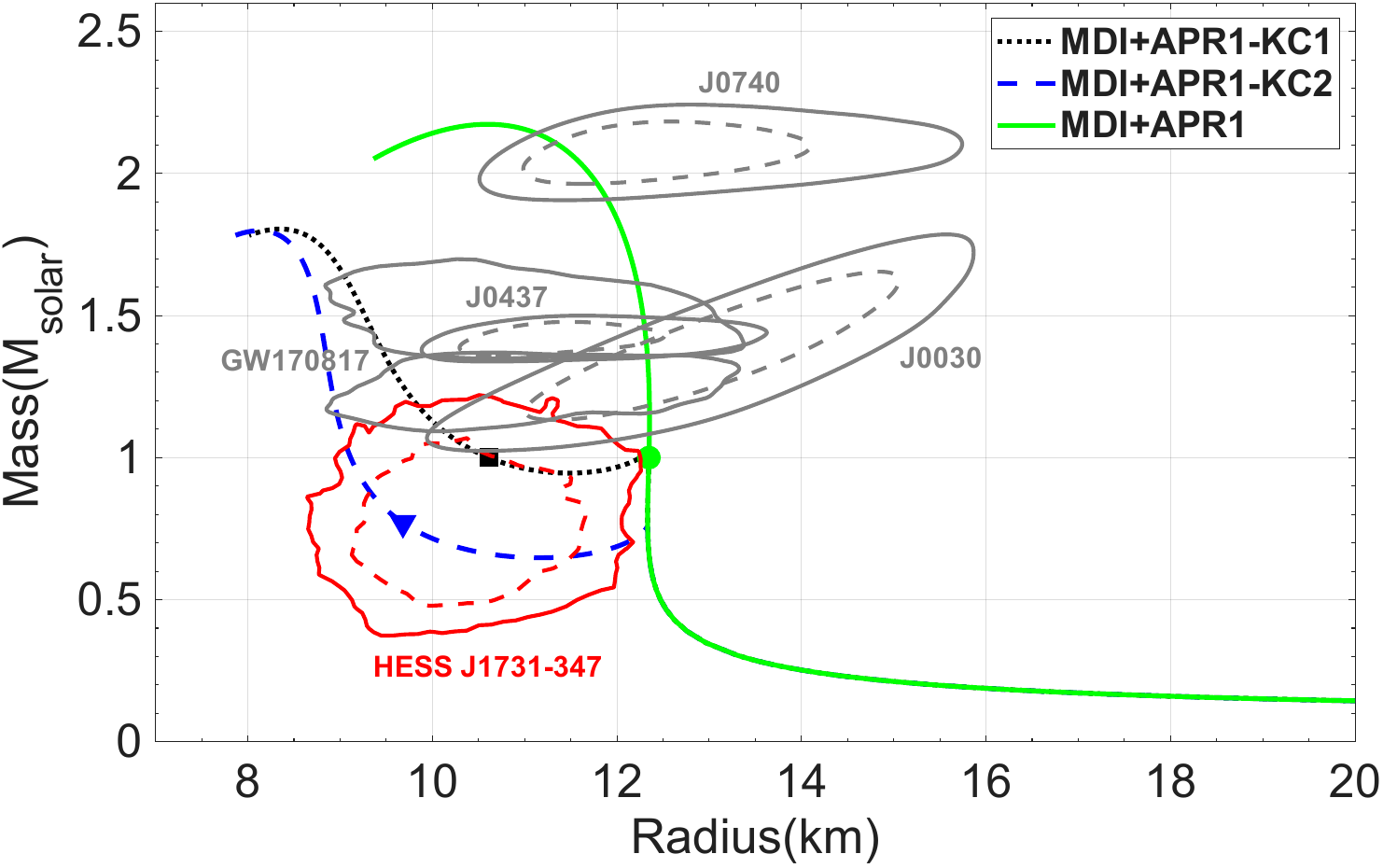}
\caption{Mass--radius relations for the EoSs considered in this work.
From top to bottom, the shaded regions represent the observational
constraints for PSR~J0740+6620~\cite{Salmi-2024},
PSR J0437--4715~\cite{Choudhury-2024},
PSR J0030+0451~\cite{Miller-2019}, and
GW170817~\cite{Abbott-2018}. The inferred mass--radius region for the
CCO in HESS J1731--347~\cite{Doroshenko-2022} is also shown. The
markers identify the stellar configurations selected for the cooling
calculations.}
\label{fig:figure1}
\end{figure}

\section{Results and Discussion}
\label{sec:Results}
\textbf{}
Figure~\ref{fig:figure1} presents the gravitational mass–radius relations predicted by the EoSs considered in this work, overlaid with the relevant observational constraints~\cite{Doroshenko-2022,Salmi-2024,Choudhury-2024,Miller-2019,Abbott-2018}. Both kaon-condensed EoSs cross the observationally inferred region for HESS J1731--347, indicating that they can reproduce the reported structural properties of the CCO. For the subsequent cooling analysis, we select two stellar configurations based on the MDI+APR1-KC1 EoS and one based on the MDI+APR1-KC2 EoS, with the latter chosen to match the inferred gravitational mass of the CCO. Although the structural properties of the purely hadronic NS are not in accordance with the mass and radius measurements of the CCO in HESS J1731--347, we examined its thermal evolution in order to highlight possible differences with the cooling of the kaon-condensed NSs. Their respective structural properties are summarized in Table~\ref{tab:2}.

\begin{figure}[t]
\centering
\includegraphics[width=\columnwidth]{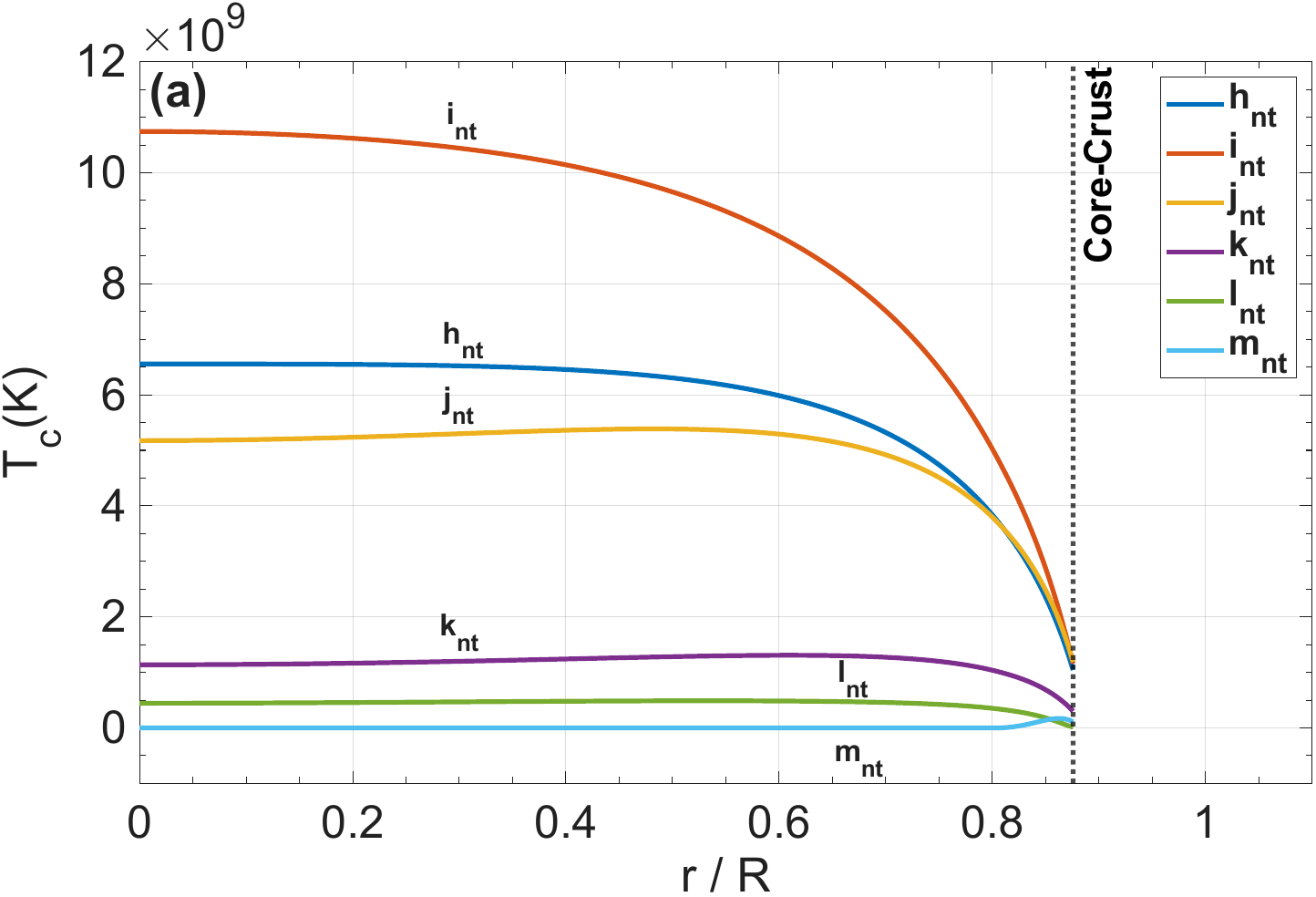}
~
\includegraphics[width=\columnwidth]{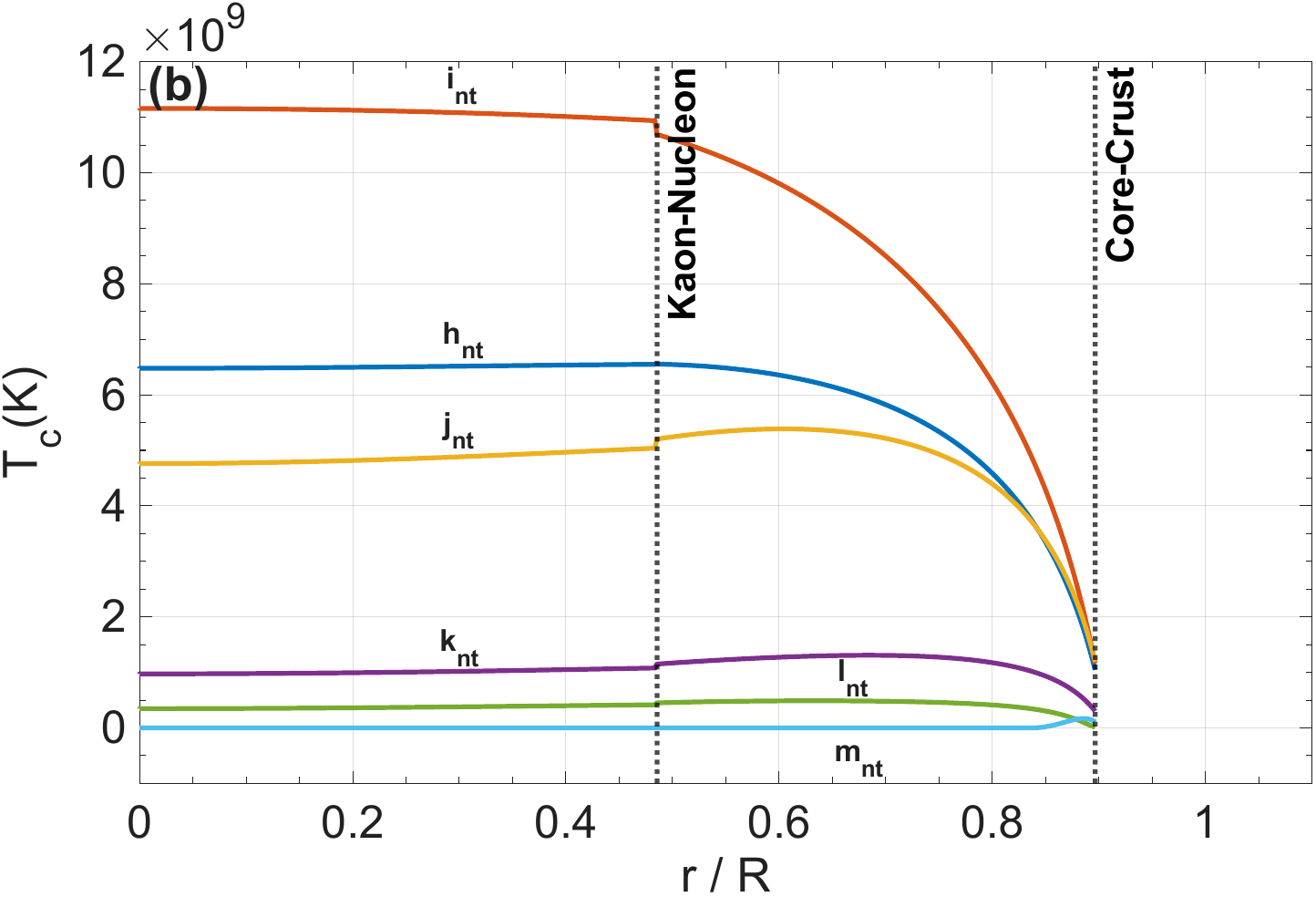}
\caption{Radial profiles of the critical temperature for the neutron ${}^{3}P_{2}$ pairing models listed in Table~1 of
Ref.~\cite{Andersson-2005}: (a) the purely hadronic MDI+APR1
configuration and (b) the kaon-condensed MDI+APR1-KC1 configuration. The labelled vertical dotted lines mark the crust--core transition in both panels and the onset of kaon condensation in panel (b). Curves of the same colour correspond to the same pairing model in both panels.}
\label{fig:2}
\end{figure}

Let us now determine the reaction profile of each model. In all stellar configurations the slow and medium neutrino emitting processes were considered in the star's core (mURCA, Bremsstrahlung, PBF)~\cite{Page-2006, Page-1996}. The fast dURCA process is not activated either in the purely hadronic star or in the hadronic regions of the kaon-condensed stars, since the corresponding triangle inequalities are not satisfied anywhere within these regions. However, the increase of proton number density in the kaon-condensed phases, triggers the activation of the dURCA process within the kaon-condensed core. In addition, the fast n-kURCA and p-kURCA processes are activated within the kaon-condensed regions. Consequently, the purely hadronic star is led into a standard cooling scenario, whereas both kaon-condensed stars undergo enhanced cooling.

\begin{figure}[t]
\centering
\includegraphics[width=0.98\columnwidth]{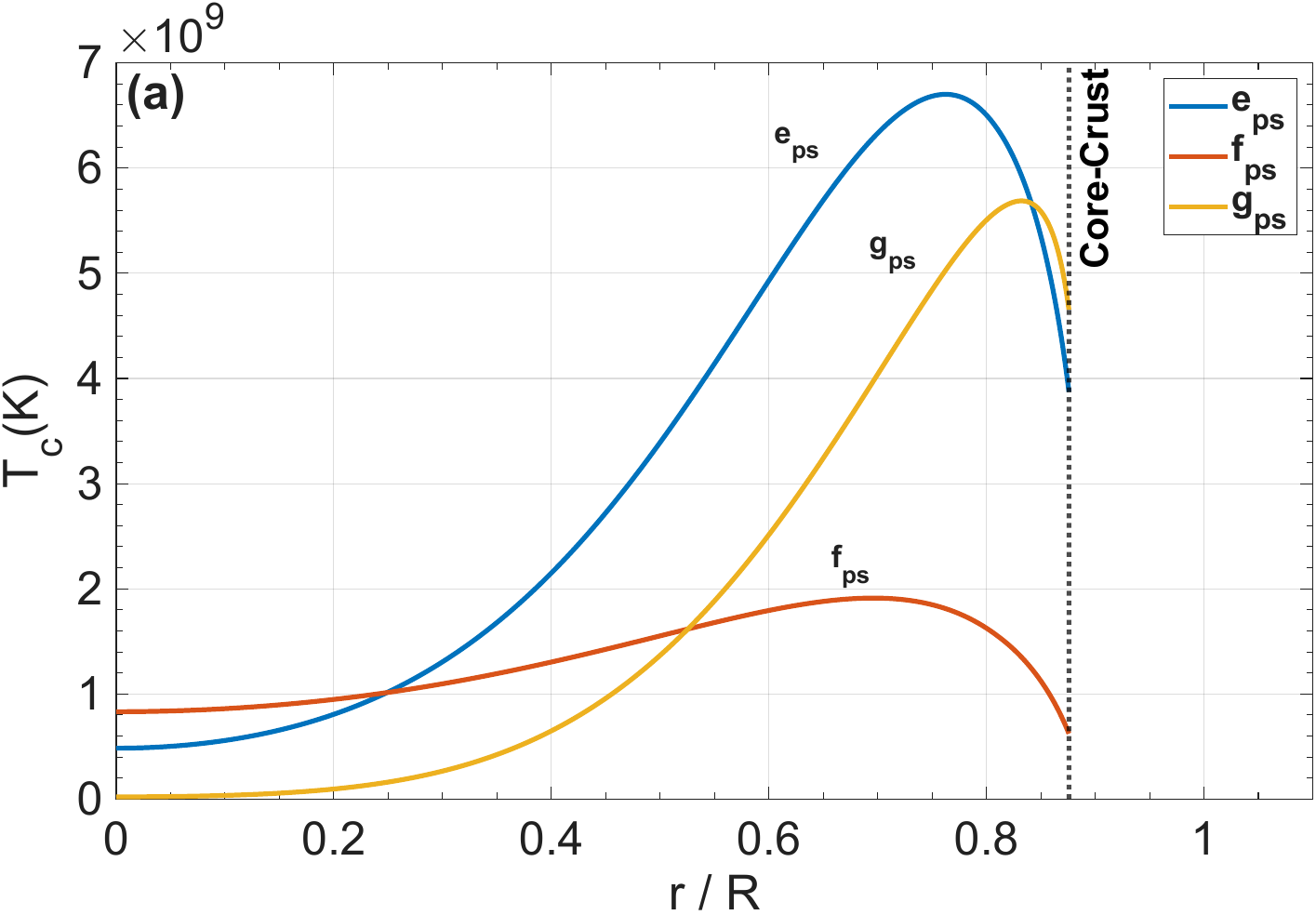}
 ~
\includegraphics[width=0.98\columnwidth]{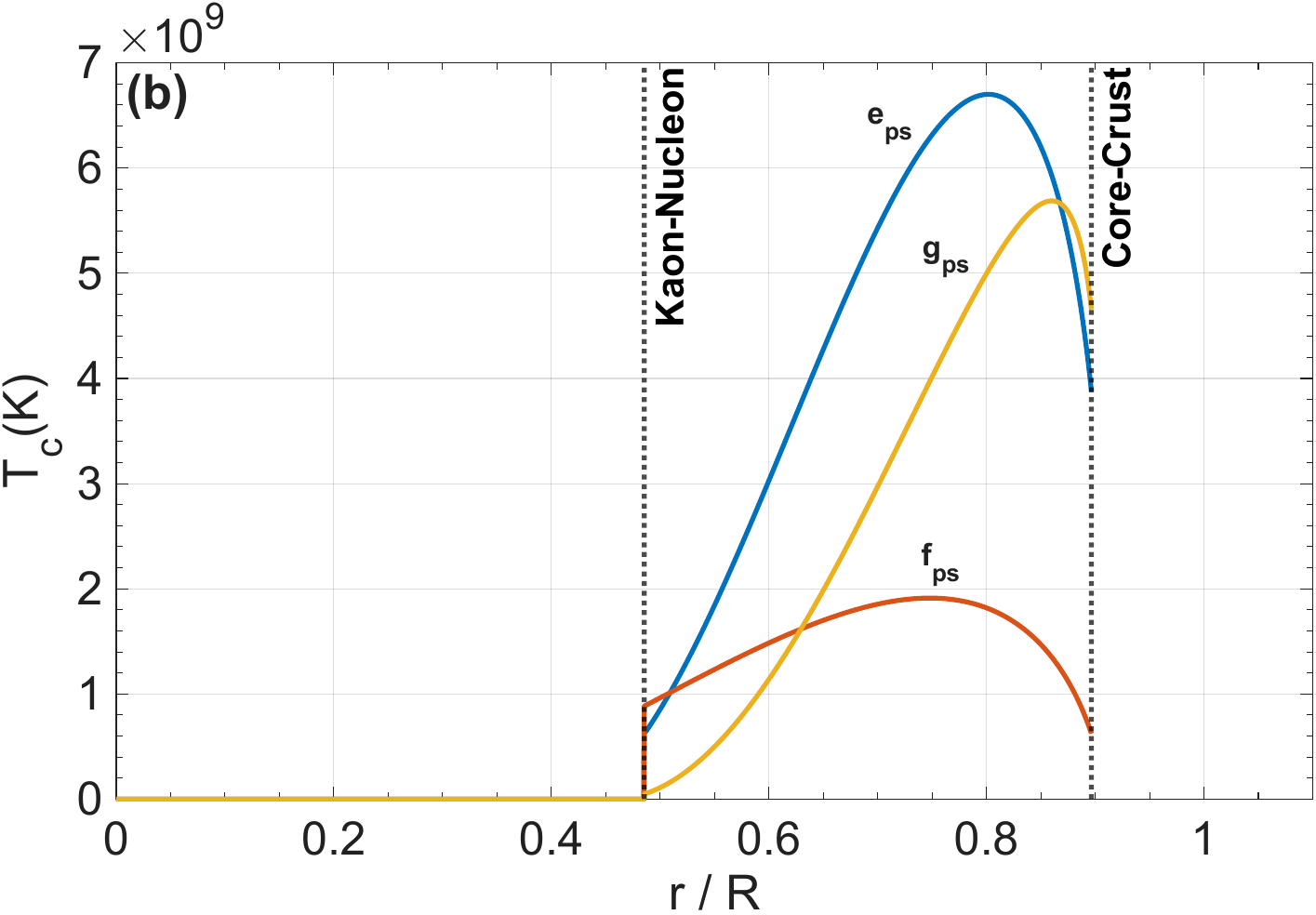}
\caption{Radial profiles of the critical temperature for the proton ${}^{1}S_{0}$ pairing models listed in Table~1 of
Ref.~\cite{Andersson-2005}: (a) the purely hadronic MDI+APR1
configuration and (b) the kaon-condensed MDI+APR1-KC1 configuration. Curves of the same colour correspond to the same pairing model in both panels.
}
\label{fig:3}
\end{figure}

\begin{table}
\caption{Structural properties of selected configurations for the considered stellar models.}
\label{tab:2}
    \begin{ruledtabular}
        \begin{tabular}{lcc}
            EoS & M ($M_\odot$) & R (km) \\
            \hline
            MDI+APR1 & 1.00 & 12.35 \\
            MDI+APR1-KC1 & 1.00  & 10.61 \\
            MDI+APR1-KC2 & 0.77  & 9.68 \\
        \end{tabular}
    \end{ruledtabular}
\end{table}

\begin{figure*}
\centering
\includegraphics[width=0.49\textwidth]{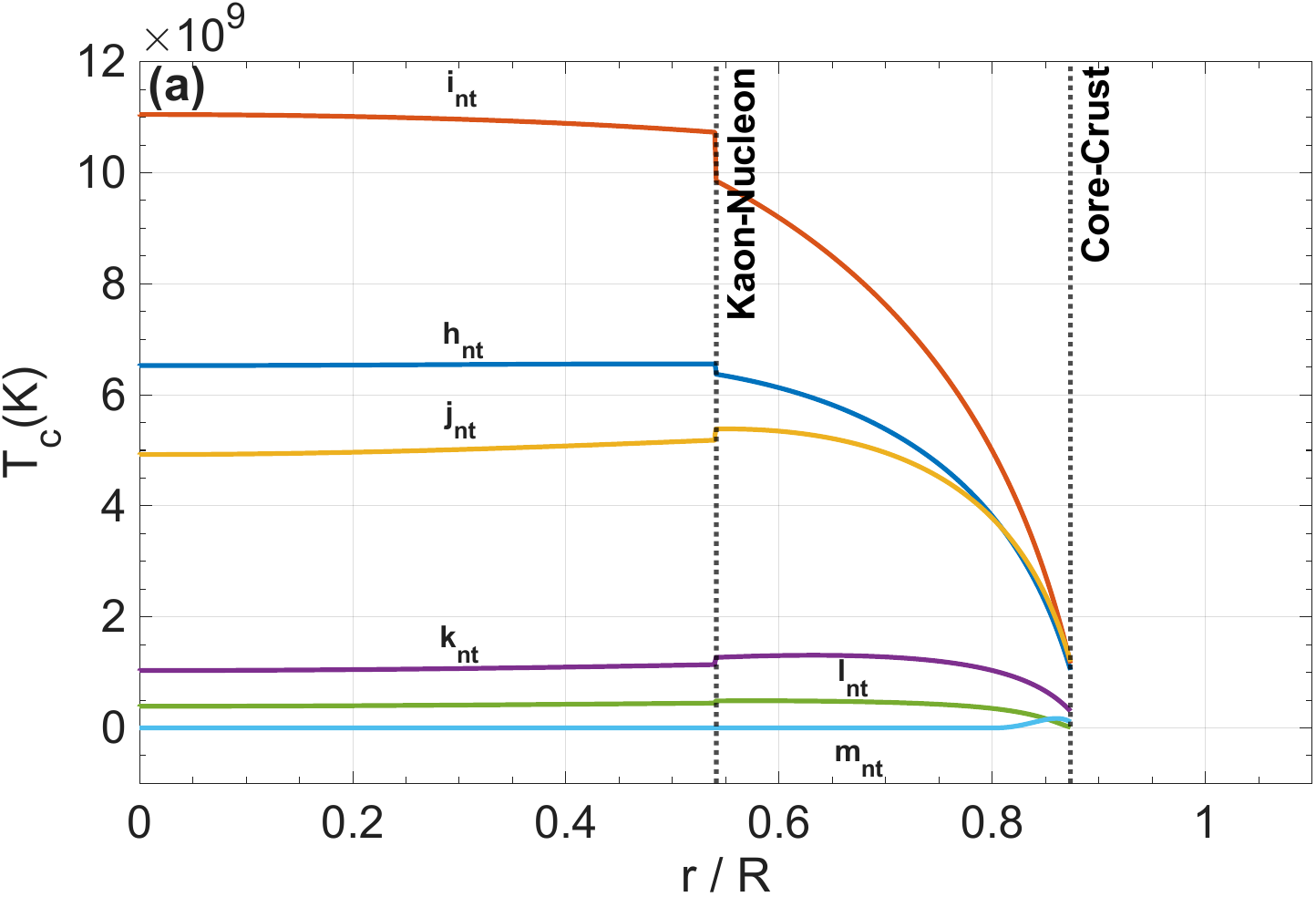}
 ~
\includegraphics[width=0.48\textwidth]{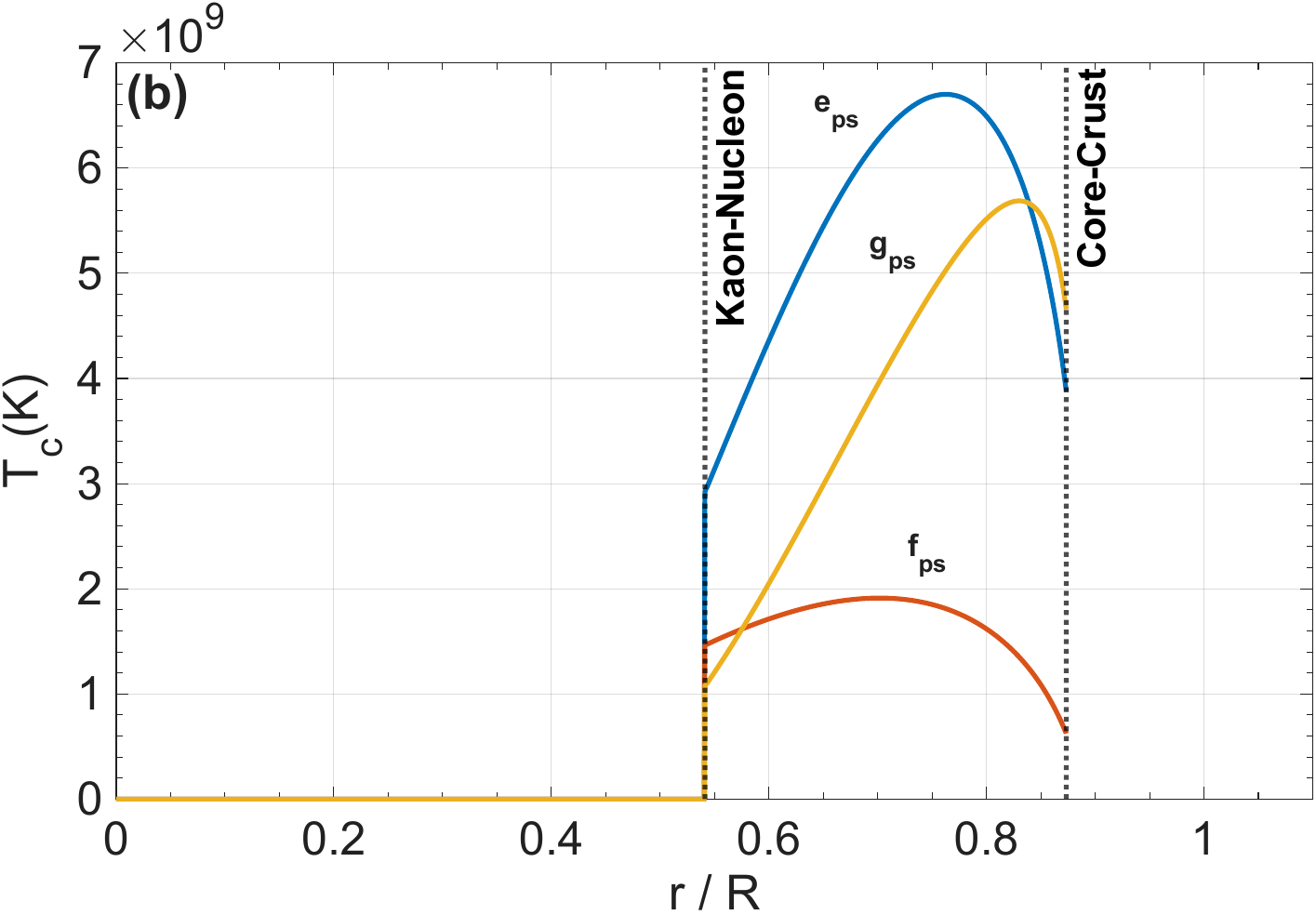}
\caption{Radial profiles of the critical temperature for the nucleon pairing models listed in Table~1 of Ref.~\cite{Andersson-2005} for the MDI+APR1-KC2 configuration: (a) ${}^{3}P_{2}$ pairing models and (b) ${}^{1}S_{0}$ pairing models. The colour schemes in the left and right panels follow Figs.~\ref{fig:2} and~\ref{fig:3}, respectively, with identical colours denoting the same neutron and proton pairing models.}
\label{fig:4}
\end{figure*}

\begin{figure*}
\centering
\includegraphics[width=0.49\textwidth]{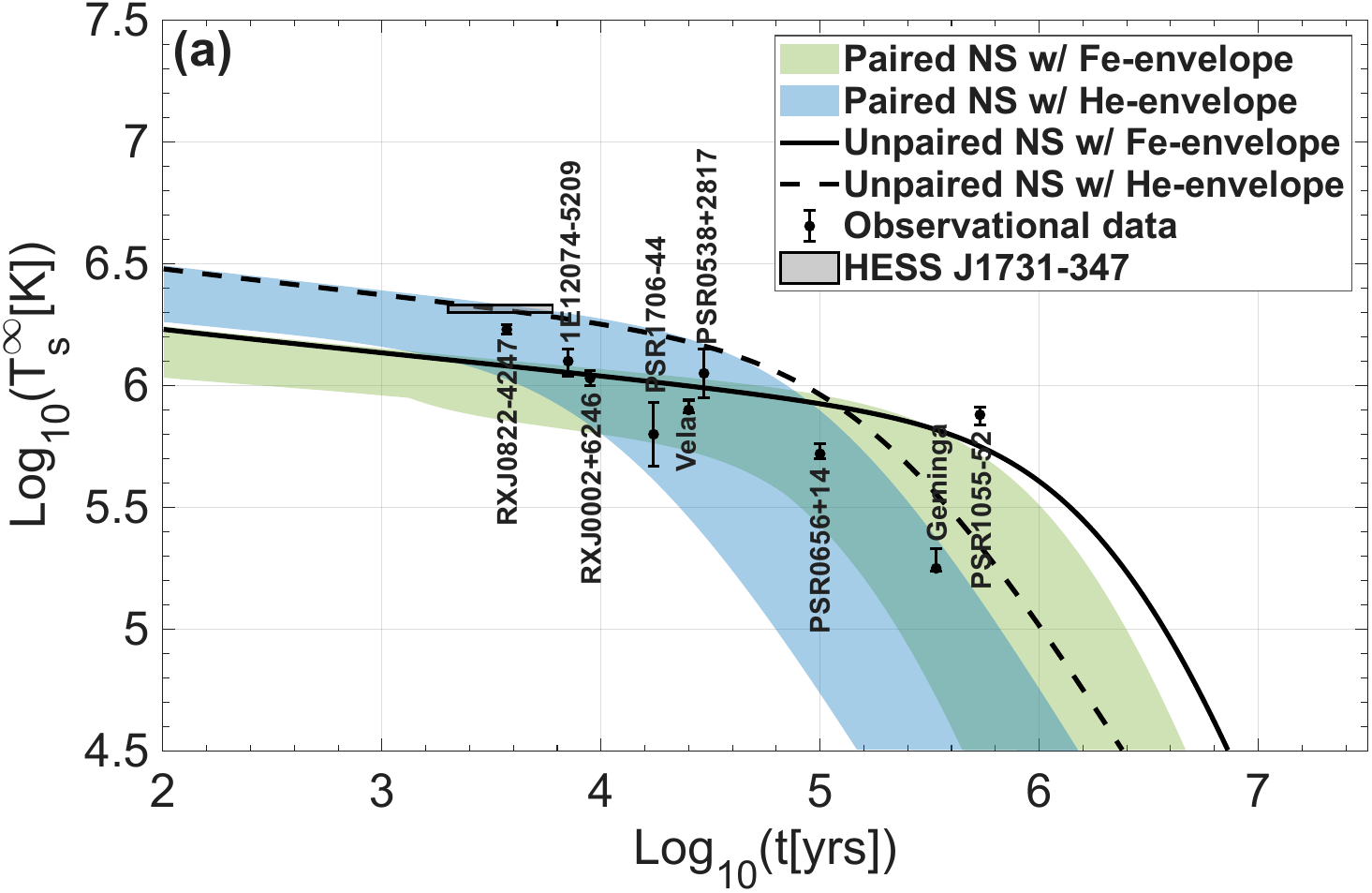}
 ~
\includegraphics[width=0.49\textwidth]{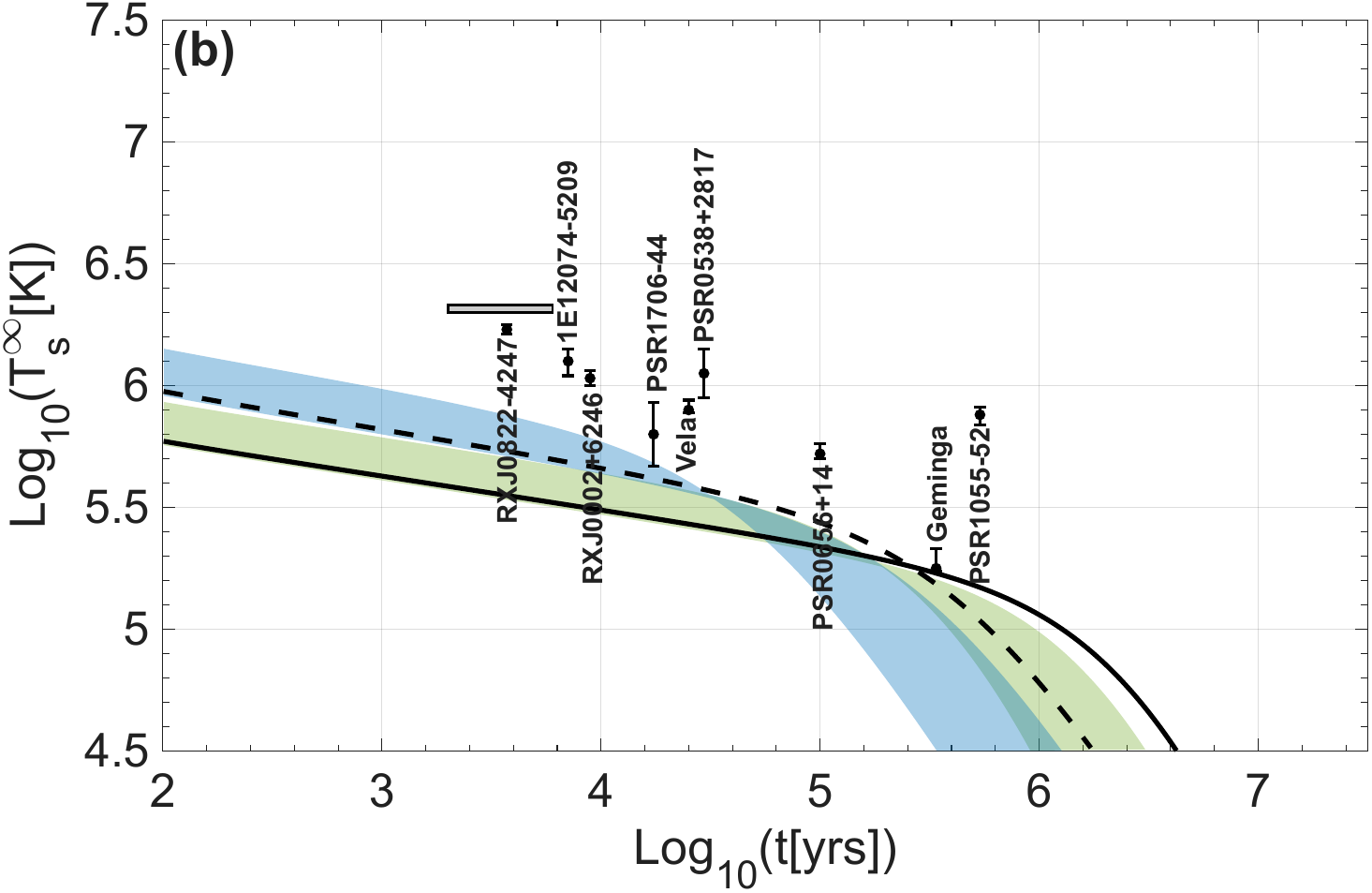}
\caption{Evolution of the redshifted surface temperature with stellar age for (a) the purely hadronic MDI+APR1 configuration and (b) the kaon-condensed MDI+APR1-KC1 configuration. In each panel, the solid black curve and the light-green shaded region correspond to the unpaired and paired cases, respectively, for a Fe-like envelope,
whereas the dashed black curve and the light-blue shaded region show the corresponding cases for an He-like envelope. The data points in Figs.~\ref{fig:5} and~\ref{fig:6} related to the observation of several pulsars were collected by the works of Refs.~\cite{Yakovlev-1999, Yakovlev-2001, Page-2004} (see also references therein). The rectangular, gray regions correspond to the redshifted, surface temperature estimation~\cite{Doroshenko-2022, Sagun-2023} with $T_s^\infty = 2.05 ^{+0.09} _{-0.06}$ MK for the CCO's estimated age $t = 2 - 6$ kyrs \cite{Horvath-2023, Acero-2015, Cui-2016, Maxted-2018}.}
\label{fig:5}
\end{figure*}

Moving on with the study of the thermal evolution, we performed 18 different cooling simulations and we extracted the corresponding curves for each stellar configuration. Each cooling simulation corresponds to a distinct combination of ${}^{1}S_{0}$ proton and ${}^{3}P_{2}$ neutron  superfluidity models, which are presented in Table 1 of Ref.~\cite{Andersson-2005}. We note that both cases of the envelope's chemical composition were considered. Figures~\ref{fig:2} and~\ref{fig:3} present the radial profiles of the neutron ($T_{c_{n}}$) and proton ($T_{c_{p}}$) critical temperatures for the ${}^{3}P_{2}$ and ${}^{1}S_{0}$ pairing models, respectively. In both figures, panels~(a) and~(b) correspond to the purely hadronic MDI+APR1 and kaon-condensed MDI+APR1-KC1 configurations, respectively. Figure~\ref{fig:4} shows the corresponding neutron and proton critical-temperature profiles for the MDI+APR1-KC2 configuration in panels~(a) and~(b), respectively.

The $T_{c_{n}}$ profile for each pairing model is similar in all three stellar configurations, as shown in Figs.~\ref{fig:2} and~\ref{fig:4}(a).
This means that the ${}^{3}P_{2}$ neutron superfluidity models are only slightly affected by the presence of kaons, extending from the core's centre to the core--crust transition density; there is an exception for the $m_{nt}$ model which vanishes near the core-crust transition density. However, this was not the case for the ${}^{1}S_{0}$ proton pairing models. As evident in Figs.~\ref{fig:3}(b) and~\ref{fig:4}(b), $T_{c_p}$ vanishes almost instantly once the kaons appear inside the star's core in both cases of the kaon-condensed stellar configurations. At the same time, the same proton pairing models reach until the core's centre of the purely hadronic star. Since ${}^{1}S_{0}$ proton pairing is confined to relatively low proton Fermi momenta~\cite{Andersson-2005}, the increase in proton number density induced by kaon condensation drives $p_{F_p}$ beyond the pairing range, causing $T_{c_p}$ to vanish within the kaon-condensed region.

\begin{figure}
    \centering
    \includegraphics[width=\columnwidth]{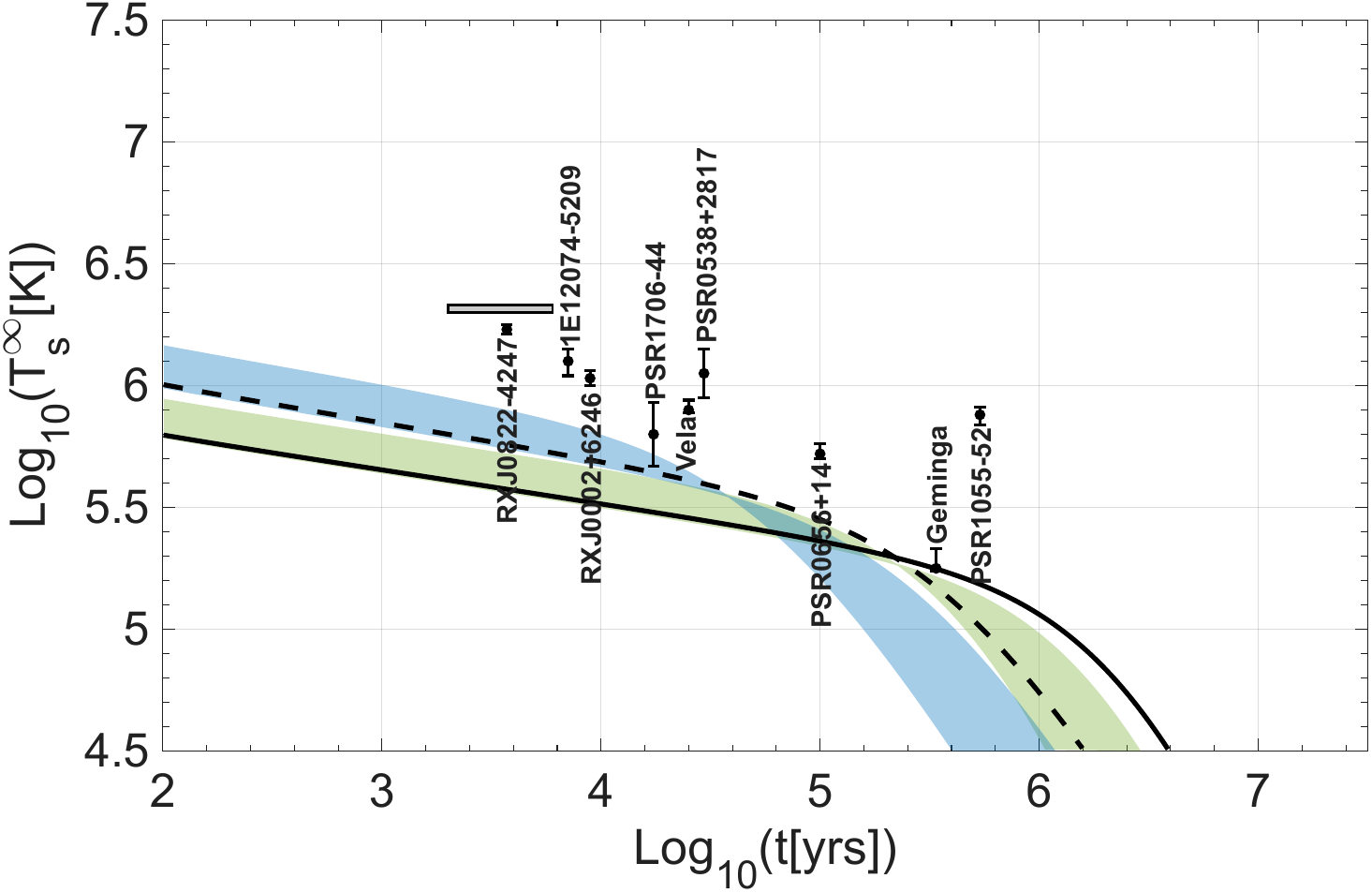}
    \caption{Evolution of the redshifted surface temperature with stellar age for the kaon-condensed MDI+APR1-KC2 configuration. The colors, line styles, and shaded regions follow the conventions of Fig.~\ref{fig:5}, denoting the same combinations of pairing and envelope composition.}
    \label{fig:6}
\end{figure}

Figure~\ref{fig:5} denotes the thermal evolution of the purely hadronic MDI+APR1 and the kaon-condensed MDI+APR1-KC1 configurations, respectively, while Fig.~\ref{fig:6} shows the corresponding results for the MDI+APR1-KC2 configuration. For the purely hadronic star, as presented in Fig.~\ref{fig:5}(a), the cases that seem to reconcile the estimated surface temperature of the CCO in the HESS J1731--347 supernova remnant are both the superfluid and non-superfluid stars with an helium-like envelope. The presence of superfluid phenomena is not necessary to predict the CCO's surface temperature, which is mainly due to the deactivation of the powerful dURCA process. Moreover, in the Fe-envelope scenario, both superfluid and non superfluid stars failed to reach the CCO's surface temperature, since heavier element envelopes, like Fe, are worse heat conductors than the ones with lighter elements, resulting in lower surface temperatures~\cite{Beznogov-2021}. Our results for the non-superfluid star cases are in excellent agreement with the work of Page et al.~\cite{Page-2004}. 

With the presence of kaons, the activation of the fast dURCA, n-kURCA, and p-kURCA processes is initiated. As shown in Fig.~\ref{fig:5}(b), the cooling simulations considering the MDI+APR1-KC1 EoS, fail to account for the CCO's properties. This fact, combined with the elimination of proton superfluidity in the kaon condensed region, forces the star to cool rapidly and deviate significantly from the CCO's surface temperature at the specific time frame. In addition, the kaon-condensed configuration matching the inferred CCO mass, presented in Fig.~\ref{fig:6}, also fails to reproduce its reported surface temperature. Although the underlying EoS and the mass of the specific model are different compared to the MDI+APR1-KC1 configuration, its thermal evolution only slightly differs from the cooling process of the previous kaon-condensed model. The suppression of proton superfluidity in the kaon-condensed region in combination with the fact that the reaction profiles of both stellar configurations are the same, lead to similar surface temperature profiles.

\section{Conclusion}
\label{sec:Conclusion}
In the present work, we examined whether a kaon-condensed core is compatible with both the structural and thermal properties inferred for the CCO in HESS J1731--347. The kaon-condensed EoSs considered here produce sufficiently compact stellar configurations to reproduce the
reported mass--radius region. However, the onset of kaon condensation strongly accelerates the thermal evolution through the activation of efficient neutrino-emission processes, driving the redshifted surface
temperature below the observational range at the estimated age of the source.

This conclusion remains unchanged for the different envelope compositions, nucleon-pairing models, and values of $a_3m_s$ considered. Although superfluidity suppresses part of the enhanced neutrino emission, it is not sufficient to recover the high surface temperature of HESS J1731--347. Therefore, within the cooling framework employed here, kaon condensation can account for the inferred compactness of the CCO, but not for its thermal state. The combined structural and cooling constraints thus do not support the interpretation of HESS J1731--347 as a kaon-condensed NS.

At this point, we should state that kaon-condensed NSs should not be excluded entirely as a possible candidate of the CCO in the HESS J1731--347 supernova remnant. Further studies of nucleon pairing in such dense, high density compact objects are necessary, which will provide a concrete and more comprehensive view on the subject. As a result, future work could examine the thermal evolution of kaon-condensed stars using alternative, microscopically motivated proton-pairing models that extend to higher baryon densities and possibly remain active throughout the stellar core. It would also be interesting to investigate the effects of pion condensation and other exotic degrees of freedom on neutron-star cooling. Finally, motivated by the work of Shrivastava et al.~\cite{Shrivastava-2026}, the thermal evolution of rapidly rotating kaon-condensed stars could be studied to explore the effect of rapid rotation on their cooling behaviour.

\section*{Acknowledgments}
This work is supported by the Croatian Science Foundation under the project Relativistic Nuclear Many-Body Theory in the Multimessenger Observation Era (HRZZ-IP-2022-10-7773). This paper was supported by the European Union – NextGenerationEU through the National Recovery and Resilience Plan 2021-2026 -- Institutional grant of University of Zagreb Faculty of Science (Nuclear Astrophysics). This research was supported by the European Union – NextGenerationEU through the National Recovery and Resilience Plan 2021–2026 Institutional grants of University of Zagreb Faculty of Science (PMF-PRESTIGE).

\section*{Author Contribution}
The study was conceptualized by D.G.N. and Ch.C.M., with Ch.C.M. providing supervision. D.G.N. and P.S.K. carried out the model validation, formal analysis, and data collection. D.G.N. additionally performed the model calculations, visualization, and figure preparation. The original manuscript was written by D.G.N., P.S.K. and Ch.C.M., and subsequently reviewed and revised by all authors. Financial support for this work was secured by P.S.K.


\bibliography{apssamp}

\end{document}